\documentstyle[eqsecnum,aps,graphicx,multicol]{revtex}
\begin{document}

\title{Crack propagation in quasicrystals}
\author{R.\,Mikulla$^1$, J.\,Stadler$^1$, F.\,Krul$^1$,
H.-R.\,Trebin$^1$ and P.\,Gumbsch$^2$}
\address{$^1$ Institut f\"ur Theoretische und Angewandte Physik der 
Universit\"at Stuttgart,\\Pfaffenwaldring 57, D-70550 Stuttgart,
Germany\\
$^2$ Max-Planck-Institut f\"ur Metallforschung, Seestr. 92, 70174
Stuttgart, Germany}

\maketitle

\begin{abstract}
  Crack propagation is studied in a two dimensional decagonal model
  quasicrystal. The simulations reveal the dominating role of highly
  coordinated atomic environments as structure intrinsic obstacles for
  both dislocation motion and crack propagation. For certain
  overloads, these obstacles and the quasiperiodic nature of the
  crystal result in a specific crack propagation mechanism: The crack
  tip emits a dislocation followed by a phason wall, along which
  the material opens up.
\end{abstract}

\begin{multicols}{2}

\narrowtext

%1
Quasicrystals are characterized by an unusual structure: They are
orientationally ordered, but with noncrystallographic, e.g., five- or
eightfold symmetry axes~\cite{janot94}. Their translational order is
not periodic, as in conventional crystals, but quasiperiodic. In the
last few years it has become possible to measure physical properties
of quasicrystals, such as mass and heat transport, plasticity, and
fracture, since high--quality, thermodynamically stable quasicrystals
could be grown in millimeter sizes~\cite{tsai96}. Because of the
peculiar structure, quasicrystals open new possibilities for studying
structure--property relations.  The mechanical behavior of a solid, in
particular, is strongly influenced by structure, because it is mainly 
governed by defects, which are related to the order.  Several studies
of the standard icosahedral quasicrystalline alloys AlMnPd and AlCuFe
have disclosed that the samples are brittle and hard at room
temperature \cite{takeu91}, but become ductile at about 80\% of the
melting temperature, and then can be deformed up to 30\% {\em without
  hardening} \cite{takeu93}.

%2
Concerning ductility it has been confirmed both by transmission
electron microscopy \cite{wollgarten93} and by computer simulations
\cite{mikulla95}, that motion of dislocations is the mechanism of
plastic deformation in quasicrystals. The existence of dislocations in
quasicrystals is due to the fact that quasiperiodic structures can be
considered as irrational cuts through a higher dimensional periodic
crystal \cite{bohsung89}. Hence the Burgers vector of a dislocation
also is of higher dimension, with a component in physical space and a
component in the complementary orthogonal space. The dislocation is
described by a ``phonon''-like displacement field $u(x)$ and a
``phason''-like displacement field $w(x)$, whose contour integrals
yield $b^{||}$ and $b^{\perp}$, respectively.

%3
Molecular dynamics simulations of a sheared quasicrystal have
revealed that close to the melting temperature the phason field is
spread out around the dislocation core, whereas at low temperatures it
is confined to a plane extending from and being bound by the
dislocation line \cite{mikulla95,mikulla95a}.  The phase of the
quasicrystal is differing by $b^{\perp}$ at the two sides of this
wall, which therefore is denoted as ``phason-wall''.

%4
In this article we demonstrate -- by employing atomistic simulations
of fracture processes -- that phason walls, being a consequence of the
quasiperiodic structure, are a crucial element in the brittle fracture
mechanism at low temperature. Furthermore, we identify atomic
clusters, inherent to the quasicrystalline structure, as obstacles and
scattering centers for the brittle crack.

%5
The study of a strongly nonlinear process like dynamical fracture in a
complex material requires simplifications. First we consider a
two dimensional system (2D) to evade the analysis of a three
dimensional crack front. Secondly we simulate a perfectly ordered
quasicystalline representative of the binary tiling system~\cite{BT},
derived from the T\"ubingen triangle tiling (TTT)~\cite{baake90}. A
large atom (L) is placed at each vertex and a small atom (S) into the
center of each large triangle of the original tiling. When the centers
of neighboring atoms of different type are connected, one obtains a
rhombic tiling, which we denote as ``bond representation'' of the
direct ``atom representation''.

\begin{figure}[htb]
  \begin{center} 
    \leavevmode
    \includegraphics[width=7.0cm]{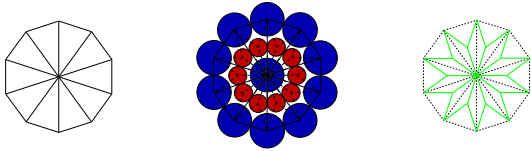}
\vspace{0.2cm}
    \caption{Ten-rings in atom (middle) and bond representation
      (right) and their origin in the triangle tiling (left).\label{F2ofKrul}}
  \end{center} 
\end{figure}

\begin{figure}[htb]
  \begin{center} 
    \leavevmode
    \includegraphics[width=7.0cm]{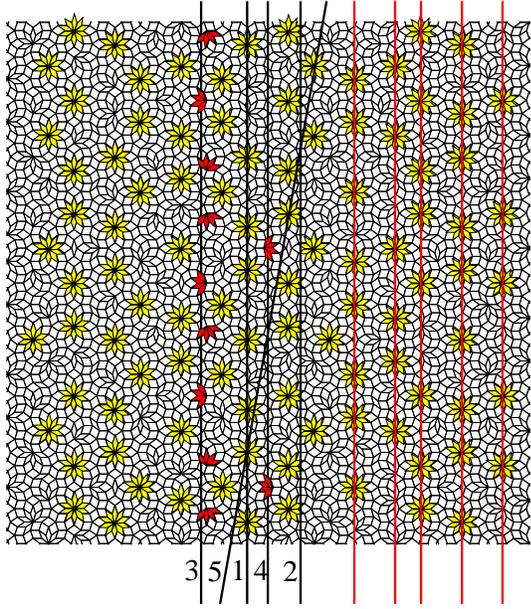}
\vspace{0.2cm}
    \caption{Binary tiling in the bond representation and special planes,
    whose surface energy is explained in the text.
\label{F3ofKrul}}
  \end{center} 
\end{figure}

%6
Qualitatively our model shares central properties with structure models
for real quasicrystals, which are (1) a decagonal diffraction pattern
inherited from the TTT~\cite{baake90} and (2) a hierarchical structure
of clusters. These are built by a first shell of small atoms and a
second shell of large atoms surrounding a central large atom
(Fig.~\ref{F2ofKrul}), and can be considered as 2D versions of Mackay
clusters that are significant for many real quasicrystals.
Furthermore, assemblies of such clusters form super clusters and
so on. The smallest clusters are situated on five families of parallel
lines, mutually rotated by $36^{\circ}$, with a small and a large
separation within each family, arranged in a Fibonacci sequence
(Fig.~\ref{F3ofKrul}).

%7
The atoms interact with Lennard-Jones potentials truncated
at 2.5 $r_{0}$. The bond lengths are chosen to be the geometrical
ones~\cite{BT}. The unit of length in our simulation $r_{0}=r_{LS}$ is
given by the length of the LS bond. The bond energies are
$\varepsilon_{0}=\varepsilon_{LS}$ and
$\varepsilon_{LL}=\varepsilon_{SS} = 0.5 \varepsilon_{0}$ to avoid
phase separation. All masses are set to 1. The unit of time is
therefore $t_{0}=r_{0}\sqrt{m/\varepsilon_{0}}$. Due to their high
coordination the decagonal clusters become the tightest bound
structural units of the quasicrystal model for all choices of
$\varepsilon_{LL}$ and $\varepsilon_{SS}$ smaller than one.

%8
Preferred cleavage planes of the 2D model quasicrystal, i.e.\ the
``easy lines'' for fracture, are those in between the large separation
of the parallel lines connecting the clusters, because a crack can
follow these lines without cutting the tightly bound ten-rings. The
surface energies \cite{surfen} of different crystallographically
equivalent, but structurally distinct lines (1-4 in
Fig.~\ref{F3ofKrul}) are lowest for those between the widely spaced
ten-rings (3,4 in Fig.~\ref{F3ofKrul}), namely 1.9~$\varepsilon_{0}/r_{0}$
as compared to 2.3~$\varepsilon_{0}/r_{0}$ for the other surfaces studied here
(1,2,5 in Fig.~\ref{F3ofKrul}).  Furthermore, they also have the
lowest ``unstable stacking fault energy'', i.e.\ increase in energy
when the two halfs are shifted relative to each other. Hence these
lines are also preferred for dislocation motion \cite{mikulla97condm}.
When a dislocation is gliding through a quasicrystal, leaving a
phason wall behind, it creates an even more preferred cleavage line,
since the phason wall is characterized by a positive defect energy.
For the shortest Burgers vector the cleavage energy along the phason
wall can be reduced by an additional 0.3~$\varepsilon_{0}/r_{0}$
\cite{mikulla97condm}.

%9
The fracture simulations are performed on a rectangular slab of
approximately 250 000 atoms, 250~$r_{0}$ high and 1.000~$r_{0}$wide.
The slab is uniformly strained by an amount $\Delta$ normal to the
long axis, and in the outermost layer of width 3$r_{0}$ all atoms are
held fixed at these positions. A precrack of length 200$r_{0}$ is
inserted from one side along an easy line by cutting the bonds and
relaxing the sample.  For the relaxation the externally applied
homogeneous strain is fixed at the Griffith value, at which the strain
energy per unit length just equals twice the surface energy. The
resulting displacement field then is scaled linearly above the
critical dilation where the crack tip starts moving. The evolution of
the system is followed by standard molecular dynamics (MD) technique
where the atoms obey Newtonian equations of motion. The system was
started at an initial temperature of T=10$^{-6}$ T$_{melt}$. The IMD
molecular dynamics package was used~\cite{stadler97}.  During crack
propagation, the crack tip bonds are broken and the strain energy
above the surface energy (the overload) is released in the form of
acoustic emission or dislocation generation.  To avoid heating of the
slab and phonon reflections from the boundaries an elliptical stadium
is created outside which the waves are damped gradually by a ramped
friction term in the equations of motion \cite{holi95}.

\begin{figure}[!htb]
  \begin{center}
    \leavevmode
    \includegraphics[width=9.0cm]{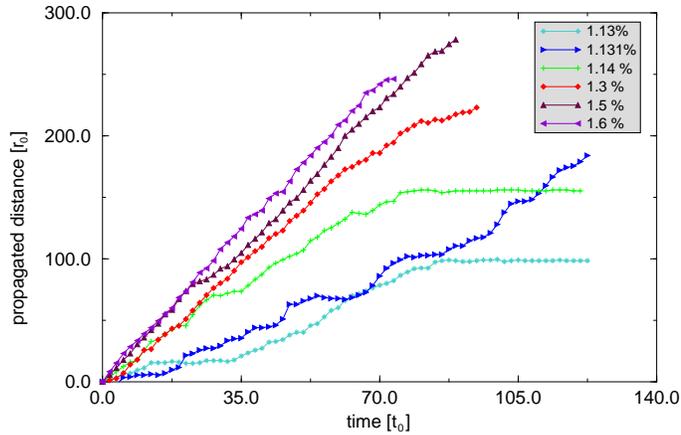}    
    \caption{Propagated distance of crack tip versus time. Depending
      on the dilation three different regimes are found: An
      intermittent ($\varepsilon$=1.13\%, 1.131\% and 1.14\%), smooth
      ($\varepsilon$=1.3\%) and unstable regime ($\varepsilon$=1.5 and
      1.6\%). \label{F3-tokyo}}
  \end{center}
\end{figure}

%10
For the phason-free precracked sample the critical strain is
$\epsilon_{c}$=1.12\%. Straight brittle cleavage along the easy
line occurred only very close to this value and for short segments. In
all other cases a rough surface is created. The propagating crack is
monitored and its current position is plotted versus time for three
different loads in Fig.~\ref{F3-tokyo}.  Depending on the load level
one can distinguish three different fracture regimes.

\begin{figure}[!htb]
  \begin{center} 
    \leavevmode
    \unitlength1cm
    \begin{picture}(5,10.2)
      \put(0.0,10.2){\includegraphics[width=3.0cm,angle=270]{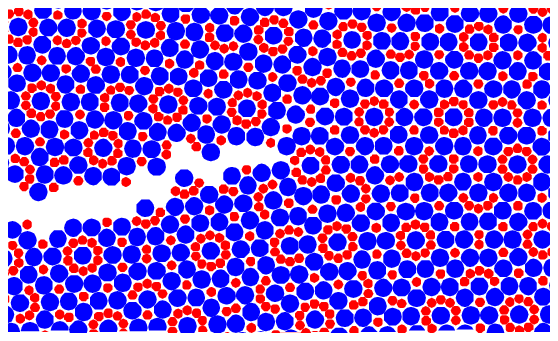}}
      \put(0.0,7.1){\includegraphics[width=3.0cm,angle=270]{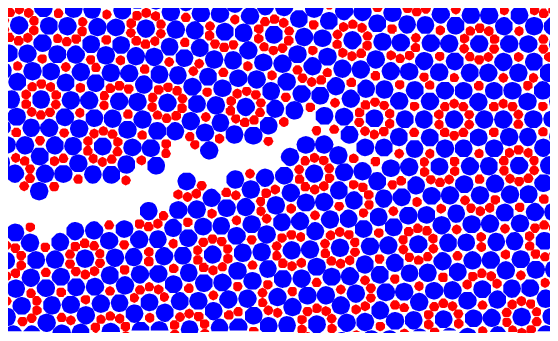}}
      \put(0.0,4.0){\includegraphics[width=3.0cm,angle=270]{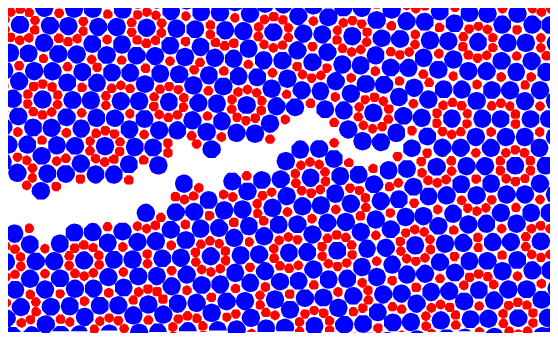}}
    \end{picture}    
    \vspace{-0.5cm}
    \caption{Crack propagation mechanism in the intermittent
      regime. Top: The crack tip has stopped. Middle: It has emitted a
      dislocation which is followed by a phason wall. Bottom: The
      quasicrystal has opened along the wall. \label{F4-tokyo}}
  \end{center} 
\end{figure}

%11
An intermittent propagation regime is found at low load levels. For
example, at a strain of 1.131\%\ the crack proceeds with constant
velocity along an easy fracture line until it hits an obstacle
(Fig.~\ref{F4-tokyo}, top), which can be a complete or an incomplete
ten-ring. It then stops and emits a dislocation along one of the easy
planes inclined by 36$^{\circ}$, along which the shear stress is high.
The dislocation moves away from the crack tip until it is stopped
itself by an obstacle(Fig.~\ref{F4-tokyo}, middle), whereupon the
entire configuration rests for some time.  The dislocation has left a
phason wall in its wake. This phason wall carries some interface
energy, and hence the cohesive strength of the material is reduced at
this wall. After some time of strain build-up the crack begins to open
along the phason wall (Fig.~\ref{F4-tokyo}, bottom). It immediately
regains its initial propagation velocity until the next obstacle is
met. Due to the tightly bound complete ten-rings, the cleavage surface
roughness is about 10~$r_{0}$.

%12
At intermediate loads, between 1.3\%\ and 1.5\%\ strain, a
regime of steady crack propagation is observed. The crack tip proceeds
with almost uniform velocity, but the appearance of the cleavage plane
is unchanged. The dislocation-emission--phason-wall cleavage mechanism
is still operates but without stopping the propagating crack.

\begin{figure}
  \begin{center} 
    \leavevmode
    \unitlength0.95cm
    \begin{picture}(9,7.0)
      \put(0.1,0.2){\includegraphics[width=3.82cm]{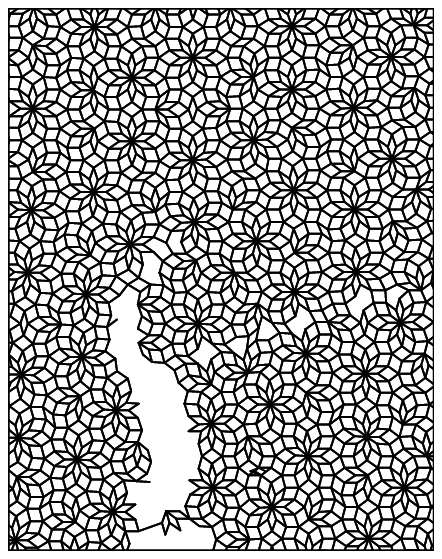}}
      \put(4.5,0.2){\includegraphics[width=3.36cm]{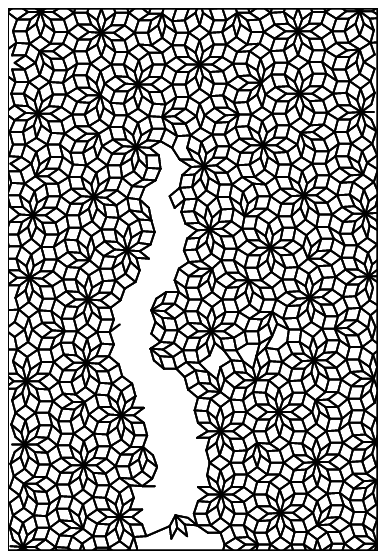}}
    \end{picture}
    \vspace{1mm}
    \caption{Virtual dislocation emission observed at a strain of
      1.5\% in the bond representation. Top: Snapshot of a simulation
      after 73.5$t_{0}$; a dislocation has been emitted at an angle of
      72$^{\circ}$. Bottom: After 77$t_{0}$ the dislocation has
      returned. \label{F5-tokyo}}
  \end{center} 
\end{figure}

%13
At even higher overloads ($\ge$ 1.5\%\ strain) we observe an onset of
crack tip instabilities, which create defects that are not bound to
the crack tip. At a strain of 1.5\% (Fig.~\ref{F5-tokyo}) we observe
dislocation emission at an angle of 72$^{\circ}$, which reduces (due to
shielding) the load on the crack tip. Arrested by an obstacle, the
dislocation moves back and annihilates at the fracture surface after
the crack tip has propagated further. Such a process, so--called
``virtual dislocation emission'', was first proposed by Brede and
Haasen to explain their fracture experiments in
silicon~\cite{brede88}. At a load of 1.7\%\ the crack bifurcates and
stops.

%14
In quasicrystals as in periodic crystals, we find a minimal velocity
for brittle crack propagation of 14\% of the shear wave velocity $v_T$
at low loads. This suggests that there exists a lower band of
forbidden crack velocities \cite{MG,GZH}. With rising load, the crack
tip speed increases to about 35\%\ of $v_T$.  This value is again
similar to the maximum crack tip speed in periodic crystals \cite{G}.
The lower than expected upper crack velocity, which linear elasticity
theory would place at the Rayleigh wave velocity \cite{Freund}
($v_{Rayleigh} \approx 0.9\,v_T$), may be explained with the
non-linearity of the atomic interaction, which has been shown to
drastically reduce crack tip velocities in periodic crystals
\cite{G,HBZ}. At high loads the upper limit on the crack velocity is
determined by the onset of crack tip instabilities: (virtual)
dislocation emission and crack branching.  It seems likely that the
instability is driven by a critical overload or a critical
accumulation of specific shear waves.

%15
While the crack tip velocity is of similar magnitude in quasicrystals
and periodic crystals, the crack propagation mode differs
significantly. At small overloads an intermittent crack propagation
mode occurs, during which the average crack velocity can be low but
microscopically separated into distinct propagation and arrest
periods. The fracture surfaces are rough, governed by cluster
inhomogeneities. Intermittency and roughness can both be attributed to
the novel dislocation-emission--phason-wall fracture mechanism.

%16
At comparable external loads the brittle fracture of perfect periodic
crystals is characterized by steady crack propagation along well
defined flat and smooth cleavage planes \cite{GZH}. In comparison to
periodic crystals, two specific structural components of the
quasicrystal are important to explain the observed behaviour. 
Firstly, the role of the ten-rings as tightly bound clusters which act
as structure intrinsic obstacles to both dislocations and cracks is
obvious. At low overloads the clusters may temporarily arrest the
crack, and even at high overloads and high crack tip speeds, the
clusters act as scattering centers for the cracks. 

%17
In perfect agreement with a previous study of dislocation mobility in
quasicrystals \cite{mikulla97condm}, we find here that the clusters
also act as obstacles to dislocation motion. As a consequence,
dislocations that have been nucleated at the crack tip cannot get
farther away than the next sufficiently strong obstacle. Since the
interaction of the crack tip and the dislocation produces a driving
force on the dislocation which decreases with distance from the crack
tip \cite{Thomson}, the strength of a cluster, measured as the
capability to arrest the dislocation, increases with distance.  Strong
local structural variations, which are a characteristic of the
quasicrystalline structure, are therefore very efficient in blocking
dislocation motion and tend to keep dislocations very close to the
crack tip. Usually this would be considered beneficial, since a
dislocation emitted from the crack tip also shields the crack tip from
the applied load, with the shielding decreasing with distance to the
crack tip \cite{Thomson}. As a consequence, one would expect a high
fracture toughness if dislocation emission occurs, while dislocations
are kept close to the crack tip. However, this is not what happens,
due to the second structural peculiarity of quasicrystals: the phason
wall left behind a moving dislocation.

%18
Since the phason wall carries excess energy, it provides a new
low--energy path for the crack and thereby effectively allows the
crack to bypass the structure-intrinsic obstacles. The existence of
the phason wall, trailing dislocations nucleated at the crack tip,
is ultimately responsible for the observed brittleness of the
quasicrystals.

%19
Although the binary model quasicrystal used in this study is very
simplistic, it reflects the most important features of real
quasicrystals, i.e.\ quasiperiodicity and clusters. It permits
conclusions relevant to real quasicrystals.  Our simulations reveal
the dominating role of highly coordinated atomic environments as
structure intrinsic obstacles for both crack and dislocation motion.
Possible toughening effects, however, are over compensated by the
embrittling nature of the phason wall. The phason wall modifies the
Griffith criterion and leads to an immediate crack opening of the
solid in the path of a single dislocation.  The existence of the
phason degree of freedom gives raise to  this special mechanism of
brittle fracture in quasicrystals.

%\newpage


\end{multicols}

\begin{thebibliography}{9999}
\bibitem{surfen} Because of the lack of periodicity, the
  surface energy of the quasicrystal must be calculated on finite
  samples. To minimize sampling errors, the surface energy is
  determined as the average energy difference per unit length between
  an intact and an (artificially) cleaved quasicrystal for several
  parts of the structure with different length.
\end{thebibliography}
\end{document}